\documentclass[conference]{IEEEtran}

\usepackage{cite}
\usepackage{amsmath,amssymb,amsfonts}
\usepackage{algorithmic}
\usepackage{graphicx}
\usepackage{textcomp}
\usepackage{xcolor}
\usepackage{booktabs}
\usepackage{balance}
\usepackage{eso-pic}
\def\BibTeX{{\rm B\kern-.05em{\sc i\kern-.025em b}\kern-.08em
    T\kern-.1667em\lower.7ex\hbox{E}\kern-.125emX}}

\begin{document}
\AddToShipoutPictureBG*{%
  \AtPageUpperLeft{%
    \put(\LenToUnit{\dimexpr(\paperwidth-\textwidth)/2\relax},\LenToUnit{-0.2in}){%
      \parbox[t]{\textwidth}{\scriptsize\centering
        Accepted to the IEEE ICDM International Workshop on AI for Personalization (WAIP 2026). To appear in the 2026 IEEE International Conference on Data Mining Workshops (ICDMW) proceedings.\\
        \copyright\,2026 IEEE. Personal use of this material is permitted. Permission from IEEE must be obtained for all other uses, in any current or future media, including reprinting/republishing this material for advertising or promotional purposes, creating new collective works, for resale or redistribution to servers or lists, or reuse of any copyrighted component of this work in other works.}}}}

\title{Improving Ads Ranking Models with Auxiliary Supervision:
Techniques and Insights}

\author{\IEEEauthorblockN{Nikhil Bhale, Apurva Pathak, SaiSujithKumarReddy MarapaReddy,
Xiao Yin, Prateek Bhansali}
\IEEEauthorblockA{\textit{Meta Platforms}\\
Menlo Park, California, USA\\
\{nbhale, apurvapathak, saisujithreddy, xiaoy, pbh\}@meta.com}
}

\maketitle

\begin{abstract}
Accurate prediction of post-click Conversion Rate (CVR) is essential for personalized ad delivery in the advertising industry. Traditional ads ranking models rely on conversion events (e.g., a user purchasing an item) as ground truth labels for training. However, these conversion events are often sparse or missing, which limits the effectiveness of model learning. The primary contributions of this paper are: (1) a case study on the use of auxiliary supervision from a diverse set of user events and metadata signals beyond traditional conversion events, and (2) advanced modeling techniques designed to maximize the value of these signals. We detail architectural innovations such as linearized labels for improved auxiliary signal precision, a split architecture to facilitate training task decomposition, and a novel Label Combination (LaCo) approach for robustly managing noisy auxiliary data. Empirical results demonstrate that these methods deliver measurable improvements in ranking models, underscoring their versatility and effectiveness across recommendation systems and related domains.
\end{abstract}


\section{Introduction}
Accurately ranking ads for user relevance is a challenge in online advertising, especially on large-scale social media platforms. These systems rely on deep neural networks to predict the probability of a click (pCTR) and the probability of a conversion given a click (pCVR) \cite{ma2018entire}.

This paper focuses on improving pCVR models, which use conversion events (such as app installs or purchases) as labels for supervised learning. However, conversion data is often sparse and unavailable for many users due to privacy choices or attribution limitations. Since only a small fraction of clicks result in conversions, modeling the user journey as a funnel, from click to conversion, highlights the need for additional signals.

\begin{figure}[!t]
  \centering
  \includegraphics[width=0.75\linewidth]{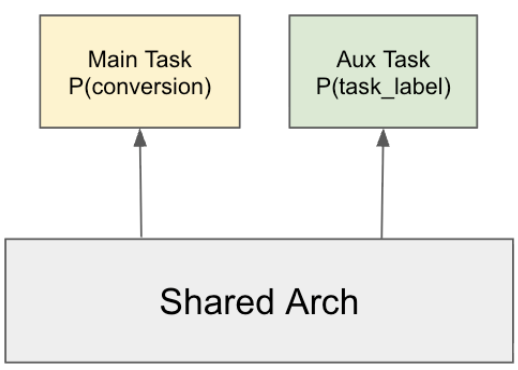}
  \caption{MTL Overview.}
  \label{fig:mtloverview}
\end{figure}

To address this, we leverage non-conversion signals and auxiliary labels that provide insight into user intent, even when conversion data is missing. Using Multi-Task Learning (MTL) \cite{multitask}, the model is trained on both the main conversion task and multiple auxiliary tasks (Fig.~\ref{fig:mtloverview}). Since auxiliary tasks are not used during inference, they do not impact online efficiency. They help the model generalize and improve performance on the main task.

Prior MTL work has focused on \textit{how} to share parameters across tasks to mitigate destructive interference --- MMoE \cite{ma2018modeling} learns task relationships through gating, PLE \cite{tangple} uses hierarchical expert separation, and others address multi-domain prediction \cite{sheng2021modelserveallstar} and data sparsity in CVR \cite{wangesdm}. For comprehensive surveys on MTL techniques, see \cite{mtlsurvey, crawshaw2020multitasklearningdeepneural}. Our work is complementary: we focus on \textit{what signals} to use as auxiliary tasks and how to handle their noise.

A related challenge is handling noise in auxiliary labels. Techniques such as DivideMix \cite{dividemix}, Tri-Training \cite{tritrain}, and Symmetric Cross-Entropy \cite{wangsce} modify training to distinguish clean from noisy samples. Our Label Combination (LaCo) approach is complementary --- it operates at the label design stage, constructing higher-quality derived labels before training begins.

Our goal is to enhance ad ranking models by incorporating these auxiliary signals, leading to better recommendations and increased revenue. The rest of the paper is organized as follows: We discuss signals that can be used as labels in Section \ref{sec:signl}, the architectural variants of auxiliary tasks, introducing the linearized labels and split architecture techniques in Section \ref{sec:archs}, and ways to handle noisy labels including our Label Combination method in Section \ref{sec:noisy}. We report experimental results in Section \ref{sec:experiments}, and the conclusions are presented in Section \ref{sec:conclude}.

\section{Signals for Auxiliary Labels}
\label{sec:signl}
The model training data consists of samples, each representing an ad impression that a user clicked on. Samples include features used for training, the main label, as well as any auxiliary labels used. The signals that can be used for labels can be broadly classified into the following categories:

\subsection{Behavioral Signals}
These are signals that represent a user's action on an ad. Examples include:
\begin{itemize}
    \item Time spent viewing an ad before clicking on it \cite{attention}.
    \item Third-party tracking events at different funnel levels, like ``add to cart'' \cite{offsite}.
    \item Time spent on the web browser before returning to the app, also known as dwell time \cite{dwelltime}.
\end{itemize}

\subsection{Semantic Signals}
These signals use semantic information (such as descriptions, images, and content) about an ad and the user's interactions. Examples include:
\begin{itemize}
    \item The category of the product browsed in a session matches the category of an ad.
    \item Certain keywords that indicate a conversion (e.g., ``confirmation'') appear in a session URL.
\end{itemize}

\subsection{Derived Signals}
Signals can also be generated using the outputs of other models, referred to as ``proxy labels''. In the proxy label approach, a model is trained on both standard features and ``privileged features'' --- information available only after the ad impression and not accessible during real-time serving. Examples of privileged features include time spent outside the app after clicking an ad or the products browsed during the session. Proxy labels are commonly incorporated through knowledge distillation \cite{distilling}, enabling the model to learn from instances where the primary label is missing. This technique allows for improved training by leveraging additional post-impression data that would otherwise be unavailable at inference time.

\subsection{Synthetic Signals}
Conversion signals are frequently scarce and often unavailable for the majority of users. To address these gaps, synthetic data can be generated from user-advertiser interaction data originating from other domains. For example, synthetic training data for a pCTR model in advertising can be generated using users' video view engagement with advertisers' content. Given that the fidelity of these auxiliary labels is typically lower than that of primary labels, they are generally utilized as auxiliary supervision rather than main supervisory signals.

\section{Architectural Variants for Auxiliary Tasks}
\label{sec:archs}

We explore various ways of using auxiliary tasks for pCVR models. In all variants, the main task and the auxiliary tasks have a shared network and a smaller, dedicated fully connected network per task. We evaluate all techniques using Normalized Entropy (NE) \cite{he2014practical}, a normalized version of binary cross-entropy loss (see Section \ref{sec:experiments} for details).

Signals that can be represented as probabilistic or boolean labels (either directly or in a normalized form) are well-suited for binary classification auxiliary tasks. For numeric labels, such as dwell time, a regression-based auxiliary task can be employed. Our experiments indicate that regression tasks are particularly effective when the primary objective of the model is also regression. Additionally, applying a logarithmic transformation to certain signals may further improve performance. In the following sections, we discuss more advanced methods for leveraging auxiliary labels.

\subsection{Linearized Labels}

For continuous or ordinal signals, linearized labels can facilitate model learning at a higher granularity than binary labels. We linearly divide the signal values into distinct buckets or classes, thereby transforming the task into a multi-class classification problem. For example, dwell time can be converted to the following classes:

\begin{itemize}
    \item More than 1 minute.
    \item Between 10 seconds and 1 minute.
    \item Between 0 and 10 seconds.
\end{itemize}
Determining appropriate bucket thresholds presents a challenge. Our findings indicate that using $n$-quantiles as bucket thresholds, with 2 to 4 buckets, yields favorable outcomes. It is important to note that bucket configurations need not be uniform across all data types. For instance, larger bucket thresholds may be suitable for E-commerce advertisements, while shorter thresholds are more appropriate for Lead generation ads, given that users typically require more time to convert on the former.

Multiple signals can also be integrated in a linearized fashion. For example, an ordered list of labels can be established based on the funnel level, such as ``conversion'', ``dwell time greater than 10 seconds'', and ``HTML content loaded'', in descending order of predictiveness of conversion. The true class is defined as the first label in the ordered list to be positive.

The NE improvements from linearized labels are reported in Table \ref{tab:results}.

\subsection{Gating}

Auxiliary labels are sometimes absent for specific data samples. For example, the dwell time signal is only available for advertisements where the ad click opens on a web browser. Training an auxiliary task on samples with missing labels by treating these missing values as negative can degrade performance. To address this issue, a gating mechanism can be implemented, ensuring that the auxiliary task is only trained on samples that meet certain predefined conditions. This approach helps maintain the integrity and effectiveness of the auxiliary task during model training.

\subsection{Split Architecture}

\begin{figure}[!t]
  \centering
  \includegraphics[width=\linewidth]{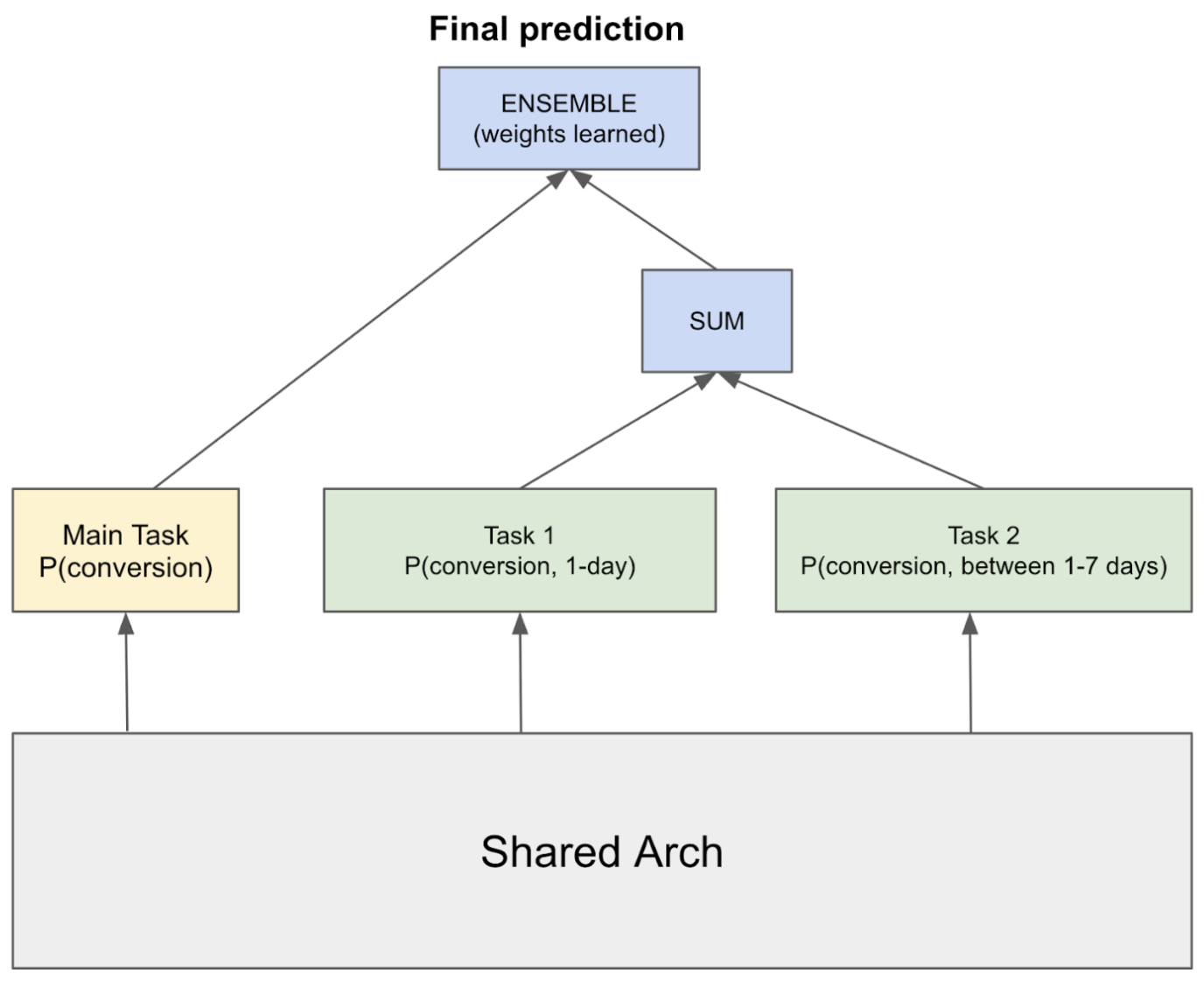}
  \caption{Split Architecture.}
  \label{fig:splitarch}
\end{figure}

The primary hypothesis of split architecture (Fig.~\ref{fig:splitarch}) is that a complex problem can be subdivided into smaller sub-problems, each addressed by dedicated experts. The outputs from these experts are then combined through ensembling to produce the final prediction. The split architecture technique decomposes the main conversion prediction task into multiple mutually exclusive and collectively exhaustive sub-tasks. For example, when modeling pCVR over a 7-day window following an ad impression, the task can be partitioned into two mutually exclusive sub-tasks: predicting pCVR within 0--1 day and within 1--7 days. The sum of these sub-task predictions is mathematically equivalent to the overall pCVR for the 0--7 day period, thereby being collectively exhaustive.

Ensembling the aggregated sub-task predictions with the main conversion prediction further improves model performance (Table \ref{tab:results}).

\section{Handling Noisy Signals}
\label{sec:noisy}
High-quality auxiliary labels are essential for enhancing model performance. Empirically, we observe that auxiliary labels closely related to the main label tend to provide greater benefits. To assess label quality, we employ precision, correlation, and recall metrics with respect to the main label.

The signals used for auxiliary labeling are inherently noisy; for instance, a long dwell time on an advertisement does not necessarily indicate a conversion. Properly accounting for such noise is crucial and can lead to further improvements in model performance.

\subsection{Sources of Noise}

The following types of noise may be present in signals:
\begin{itemize}
    \item \textbf{Labeling errors}: These are errors that come from logging inconsistencies. The mitigation of such errors can be achieved through outlier filtering.
    \item \textbf{Missing data}: Labels may not be available for all rows. For example, the dwell time signal cannot be used for clicks that do not open in a browser.
    \item \textbf{Weak labels}: The signals that are more upper funnel (e.g., HTML content loaded) will naturally be less predictive of conversions. Precision of signals improves as we go down the funnel, and recall decreases. Derived signals can help mitigate this type of noise.
\end{itemize}

\subsection{Label Combination (LaCo)}

When using multiple signals, having a unique auxiliary task for each label can sometimes degrade performance due to destructive interference between tasks. The Label Combination technique (LaCo) addresses this by selectively identifying and combining the most effective signals into a single derived label with reduced noise.

\subsubsection{Method}
LaCo employs the F-Beta score to guide label selection, which is the weighted harmonic mean of precision and recall:
\begin{equation}
F_\beta = (1 + \beta^2) \cdot \frac{\text{precision} \cdot \text{recall}}{(\beta^2 \cdot \text{precision}) + \text{recall}}\label{eq:fbeta}
\end{equation}
where precision and recall are measured with respect to the main conversion label. By tuning $\beta$, the method controls the tradeoff between precision (label quality) and recall (label coverage). The search procedure evaluates all combinations of available signals using logical operators (AND, OR, NOT OR, NOT AND, XOR) and selects the combination maximizing the F-Beta score. This approach abstracts domain-specific information by assuming that essential characteristics are adequately captured within the precision and recall metrics. In practice, with $n \leq 10$ signals per model, exhaustive search is tractable.

\subsubsection{Tuning $\beta$}
We evaluated $\beta$ values ranging from 0.001 to 10 on a conversion prediction model. The best result was achieved at $\beta = 0.02$, yielding high precision (91\%) with moderate recall. Values of $\beta \geq 1$ consistently degraded performance, confirming that precision is more critical than recall for auxiliary label quality. Our strategy is to first ensure high precision (around 90\%) and then maximize recall within that constraint. It is worth noting that the optimal $\beta$ varies across models depending on the specific data distribution and training pipeline. Overall NE improvements are reported in Table \ref{tab:results}.

\subsection{Noisy Label Training Techniques}

Certain techniques attempt to address the issue of noisy labels by directly modifying the architectural and training methodologies. We highlight two techniques found to be beneficial.

\textbf{DivideMix} \cite{dividemix} helps models learn from noisy labels by splitting training data into ``clean'' and ``noisy'' sets based on how well the model predicts each sample. Clean samples are used for regular training, while noisy ones are handled with special techniques like MixMatch and Mixup, which blend data and labels to improve learning.

\textbf{Tri-Training} \cite{tritrain} uses three models that teach each other. Each model learns from data labeled by the other two, but only when they agree, which helps filter out bad labels. This method is especially useful when true labels are rare or noisy. In our pCVR models, Tri-Training has further improved results by making training more robust to label errors.

The NE improvements from these techniques are reported in Table \ref{tab:results}.

\section{Experiments}
\label{sec:experiments}

We evaluate the proposed techniques on pCVR models used in ads ranking.

\subsection{Setup}

Our baseline is a pCVR model that serves as the core conversion prediction component in a large-scale ad ranking system. The model is a deep sparse neural network with a shared-bottom multi-task architecture using hard parameter sharing. Sparse embedding layers and shared dense layers form a common backbone, with task-specific fully connected heads on top. Each auxiliary task has a dedicated loss weight, optimized via hyperparameter tuning. It is trained on ad click data, processing billions of examples daily with a daily refresh cycle, to predict the probability of a post-click conversion event (e.g., purchase, app install) within a specified attribution window.

Our training data consists of ad impressions where the user clicked on the ad. Each row contains user and ad features, the main conversion label, and auxiliary signal labels. Approximately 50\% of training samples have missing conversion labels. We use Normalized Entropy (NE) \cite{he2014practical} as our primary offline evaluation metric, where lower NE indicates better model performance.

\subsection{Results}

Table \ref{tab:results} summarizes the offline NE improvements of each technique over the baseline pCVR model.

\begin{table}[!t]
\renewcommand{\arraystretch}{1.3}
\caption{Offline NE improvements over baseline {\upshape pCVR} model (baseline NE = 0.54). Lower NE is better.}
\label{tab:results}
\centering
\begin{tabular}{lcc}
\toprule
\textbf{Technique} & \textbf{NE} & \textbf{Relative Improvement} \\
\midrule
Baseline & 0.5400 & --- \\
+ Linearized Labels & 0.5392 & $-0.15\%$ \\
+ Split Architecture & 0.5393 & $-0.13\%$ \\
+ LaCo & 0.5387 & $-0.25\%$ \\
+ Noisy Label Training & 0.5395 & $-0.10\%$ \\
\bottomrule
\end{tabular}
\end{table}

LaCo provides the largest individual gain, confirming that systematic label combination is more effective than using individual noisy signals as separate auxiliary tasks. The linearized labels and split architecture techniques offer complementary improvements by enabling finer-grained supervision and task decomposition, respectively.

\subsection{Online Impact}

These techniques were deployed as part of a broader set of improvements to our pCVR ranking system. The combined launch, which included the techniques described in this paper alongside other improvements, resulted in a statistically significant improvement in advertiser conversion volume in online A/B tests. While we cannot isolate the individual online contribution of each technique, offline NE improvements of this magnitude have historically correlated with measurable online gains in our system.

\section{Conclusion}
\label{sec:conclude}

This paper presented several techniques for leveraging non-objective signals via multi-task learning to improve pCVR models in online advertising. By incorporating behavioral, semantic, derived, and synthetic signals, and by employing architectural variants like linearized labels and split architecture, we achieved notable gains in model performance. Furthermore, we addressed the challenge of noisy auxiliary data through techniques such as Label Combination (LaCo). These approaches have demonstrated consistent effectiveness. Each technique individually yielded NE improvements ranging from 0.1\% to 0.25\%. Such gains are highly significant within the context of online advertising at scale, underscoring their broad applicability and substantial impact. Although the primary focus is on pCVR models, these techniques possess extensive utility across recommendation systems and other relevant domains. We hope this paper serves as a practical guide for practitioners looking to leverage auxiliary supervision in their own ranking and recommendation systems.

\section*{Acknowledgment}

To Ali Mashhoori, Tom Nguyen, Devansh Goenka, Sammy Bald, and Yao Yu for their contributions to implementation and experimentation. To Pankaj Sitpure, Jon Hickman, and Pramod Gurunath for their support and feedback. To our colleagues at Meta for their contributions that enabled this work.

\balance
\bibliographystyle{IEEEtran}
\bibliography{paper-ref-ieee}

\end{document}